\documentclass[preprint,authoryear,12pt]{elsarticle}

\usepackage{graphicx}

\usepackage{amssymb}
\usepackage{amsthm}

\usepackage{lineno}

\usepackage{amsmath}
\usepackage{amsfonts}
\usepackage{bbm}
\usepackage{bm}
\usepackage{mathrsfs}
\usepackage{booktabs}
\usepackage{multirow}
\usepackage{enumerate}
\usepackage{setspace}

\usepackage{lscape}

\usepackage[table]{xcolor}

\newcommand{\E}{\mathsf{E}}
\newcommand{\var}{\mathsf{Var}\,}
\newcommand{\cov}{\mathsf{Cov}}

\newcommand{\mse}{\mathsf{MSE}}
\renewcommand{\P}{\mathsf{P}}

\newcommand{\dif}{\partial}
\newcommand{\QIC}{\mathrm{QIC}}
\newcommand{\CIC}{\mathrm{CIC}}

\newcommand{\bA}{\mathbf{A}}
\newcommand{\bB}{\mathbf{B}}
\newcommand{\bC}{\mathbf{C}}

\newcommand{\bX}{\mathbf{X}}

\newcommand{\bz}{\mathbf{z}}

\newcommand{\bV}{\mathbf{V}}

\newcommand{\bS}{\mathbf{S}}

\newcommand{\bI}{\mathbf{I}}

\newcommand{\bH}{\mathbf{H}}
\newcommand{\bu}{\mathbf{u}}

\newcommand{\be}{\mathbf{e}}

\newcommand{\br}{\mathbf{r}}
\newcommand{\bD}{\mathbf{D}}
\newcommand{\zero}{\mathbf{0}}

\newcommand{\bvartheta}{\bm{\vartheta}}

\newcommand{\bth}{\bm{\theta}}

\newcommand{\bSigma}{\bm{\Sigma}}

\newcommand{\bmu}{\bm{\mu}}

\newcommand{\wh}[1]{\widehat{#1}}

\theoremstyle{plain}

\theoremstyle{definition}

\theoremstyle{remark}

\journal{\color{white} Insurance: Mathematics and Economics}

\begin{document}

\begin{frontmatter}



\title{Modeling Dependencies in Claims Reserving with GEE}


\author[mff]{\v{S}\'{a}rka Hudecov\'{a}} \ead{hudecova@karlin.mff.cuni.cz}

\author[mff]{Michal Pe\v{s}ta\corref{cor1}} \ead{pesta@karlin.mff.cuni.cz}

\cortext[cor1]{Corresponding author, \emph{tel:} (+420) 221 913 400, \emph{fax:} (+420) 222 323 316}

\address[mff]{Charles University in Prague, Faculty of Mathematics and Physics, Department of Probability and Mathematical Statistics, Sokolovsk\'{a} 83, 186~75 Prague~8, Czech Republic}

\begin{abstract}
A~common approach to the claims reserving problem is based on generalized linear models (GLM). Within this framework, the claims in different origin and development years are assumed to be independent variables. If this assumption is violated, the classical techniques may provide incorrect predictions of the claims reserves or even misleading estimates of the prediction error.

In this article, the application of generalized
estimating equations (GEE) for estimation of the claims reserves is shown.
Claim triangles are handled as panel data, where claim amounts within the same accident year are dependent. Since the GEE allow to incorporate dependencies, various correlation structures are introduced and some practical recommendations are given.

Model selection criteria within the GEE reserving method are proposed. Moreover, an estimate for the mean square error of prediction for the claims reserves is derived in a~nonstandard way and its advantages are discussed. Real data examples are provided as an illustration of the potential benefits of the presented approach.
\end{abstract}


\end{frontmatter}



\section{Introduction}
Claims reserving is a~classical problem in general insurance. A~number of various methods has been invented, see \cite{england2002} or \cite{wutrich_kniha} for an~overview. Among them, generalized linear models (GLM) have become a~common statistical tool for modeling
actuarial data.

All the classical approaches are based on the assumption that the claim amounts in different years are independent variables. However, this assumption can be sometimes unrealistic or at least questionable. It has been pointed out that methods, which enable \emph{modeling the dependencies}, are needed, cf.~\cite{antonio} or \cite{antonio2}. The mentioned papers suggest the generalized linear mixed models (GLMM) to handle the possible dependence among the incremental claims in successive development years. This approach extends the classical GLM and is frequently used in panel (longitudinal) data analyses. In this paper, we present the use of another possible extension of GLM, namely the \emph{generalized estimating equations} (GEE) method. 

GEE were introduced by \cite{liang} as a~method for estimating model parameters if the independence assumption is violated.  The primary interest of the analysis is to model the marginal expectation of the response variable given the covariates. In contrast to GLMM, this method does not explicitly model the correlation structure. The associations are treated as nuisance parameters and modeled using so called ``working correlation matrices''. The method yields consistent and asymptotically normal parameter estimates even though the correlation structure is misspecified~\citep[Sec.~5.2]{gee2}. 
In addition, \emph{no additional distributional assumptions} are required, compared to a~specific probability distribution for the outcome in the GLM (or even GLMM) framework. The GEE solely assume that the distribution belongs to the exponential family.
   
The claims reserving notation is introduced in Section~\ref{sec_claims}. In Section~\ref{sec_gee}, the principles of the GEE together with their application within claims reserving are explained. Some model selection criteria, which can be used in the GEE setup, are presented in Section~\ref{sec_model}. The fact, that the observations of a~common accident year are correlated, is taken into account in the derivation of the mean square error (MSE) of prediction for the claims reserves. Moreover, Section~\ref{sec_mse} elaborates a~non-traditional way of estimating the MSE of prediction. Finally, two real data examples are presented in Section~\ref{sec_data}. The results illustrate potential benefits of the GEE method and the suggested estimate of the MSE of predicted claims.
   

\section{Claims Reserving Notation}\label{sec_claims}
We introduce the classical claims reserving notation and terminology. Outstanding loss liabilities are structured in so-called claims development triangles, see Table~\ref{tab:run-off}. Let us denote $X_{i,j}$ all the claim amounts in development year $j$ with accident year $i$. Therefore, $X_{i,j}$ stands for the \emph{incremental claims} in accident year $i$ made in accounting year $i+j$. The current year is $n$, which corresponds to the most recent accident year and development period as well. That is, our data history consists of right-angled isosceles triangle $\{X_{i,j}\}$, where $i=1,\ldots,n$ and $j=1,\ldots,n+1-i$.
\begin{table}[!ht]
\begin{center}
\begin{tabular}{c|>{\centering}p{1.2cm}<{\centering}>{\centering}p{1.2cm}<{\centering}>{\centering}p{1.2cm}<{\centering}>{\centering}p{1.2cm}<{\centering}>{\centering}p{1.2cm}<{\centering}>{\centering}p{1.2cm}p{1.2cm}<{\centering}}
\hline
Accident & \multicolumn{7}{c}{Development year $j$} \\
\cline{2-8}
year $i$ & \cellcolor[gray]{.6} $1$ & \cellcolor[gray]{.6} $2$ & \cellcolor[gray]{.6} & \cellcolor[gray]{.6} $\cdots$ & \cellcolor[gray]{.6} & \cellcolor[gray]{.6} $n-1$ & \cellcolor[gray]{.6} $n$ \\
\hline
\cellcolor[gray]{.6} $1$ & \cellcolor[gray]{.8} $X_{1,1}$ & \cellcolor[gray]{.8} $X_{1,2}$ & \cellcolor[gray]{.8} & \cellcolor[gray]{.8} $\cdots$ & \cellcolor[gray]{.8} & \cellcolor[gray]{.8} $X_{1,n-1}$ & \cellcolor[gray]{.8} $X_{1,n}$ \\
\cellcolor[gray]{.6} $2$ & \cellcolor[gray]{.8} $X_{2,1}$ & \cellcolor[gray]{.8} $X_{2,2}$ & \cellcolor[gray]{.8} & \cellcolor[gray]{.8} $\cdots$ & \cellcolor[gray]{.8} & \cellcolor[gray]{.8} $X_{2,n-1}$ & \\
\cellcolor[gray]{.6} & \cellcolor[gray]{.8} & \cellcolor[gray]{.8} & \cellcolor[gray]{.8} $\ddots$ & \cellcolor[gray]{.8} & \cellcolor[gray]{.8} && \\
\cellcolor[gray]{.6} $\vdots$ & \cellcolor[gray]{.8} $\vdots$ & \cellcolor[gray]{.8} $\vdots$ & \cellcolor[gray]{.8} & \cellcolor[gray]{.8} $X_{i,n+1-i}$ &&& \\
\cellcolor[gray]{.6} & \cellcolor[gray]{.8} & \cellcolor[gray]{.8} & \cellcolor[gray]{.8} &&&& \\
\cellcolor[gray]{.6} $n-1$ & \cellcolor[gray]{.8} $X_{n-1,1}$ & \cellcolor[gray]{.8} $X_{n-1,2}$ &&&&& \\
\cellcolor[gray]{.6} $n$ & \cellcolor[gray]{.8} $X_{n,1}$ &&&&&& \\
\hline
\end{tabular}
\end{center}
\caption{Run-off triangle for incremental claim amounts $X_{i,j}$.}
\label{tab:run-off}
\end{table}

Suppose that $Y_{i,j}$ are \emph{cumulative payments} or \emph{cumulative claims} in origin year $i$ after $j$ development periods, i.e., $Y_{i,j}=\sum_{k=1}^{j}X_{i,k}$. Hence, $Y_{i,j}$ is a~random variable of which we have an observation if $i+j<n+1$ (a~run-off triangle). The aim is to estimate the ultimate claims amount $Y_{i,n}$ and the outstanding \emph{claims reserve} $R_i^{(n)}=Y_{i,n}-Y_{i,n+1-i}$ for all $i=2,\dots,n$.

\section{Generalized Estimating Equations}\label{sec_gee}
Run-off triangles are comprised by observations which are ordered in time. It is therefore natural to suspect the observations to be correlated. Probably the most natural approach is to assume that the observations of a common accident year are correlated---they form a~\emph{cluster}. On the other hand, observations of different accident years are supposed to be independent. This assumption is similar to those of the Mack's chain ladder model, cf.~\cite{mack}.

Consider that the incremental claims for accident year $i\in\{1,\ldots,n\}$ create an~$(n-i+1)\times 1$ vector $\bX_i=[X_{i,1},\ldots,X_{i,n-i+1}]^{\top}$. It is assumed that the vectors $\bX_1,\dots,\bX_n$ are independent, but the components of $\bX_i$ are allowed to be correlated. Hence, the claim triangle can be considered as a~specific type of \emph{panel data} with accident year (row) clusters.

In the next sections we explain the main principles of GEE and give some recommendation for the use within the claims reserving. We refer to~\cite{gee1} and~\cite{gee2} for further reading on this topic.
%
%

\subsection{Three Pillars of GEE}\label{sec_gee_pillars}
Denote the expectation of $\bX_i$ as
\[
\E\bX_i=\bmu_i=[\mu_{i,1},\ldots,\mu_{i,n-i+1}]^{\top}.
\]
Suppose that accident year $i$ and development year $j$ influence the expectation of claim amount via so-called \emph{link function} $g$ in the following manner:
\begin{equation}\label{muij}
\mu_{i,j}=g^{-1}(\eta_{i,j})=g^{-1}(\bz_{i,j}^{\top}\bth),
\end{equation}
where $g^{-1}$ is the inverse of scalar link function $g$ and $\bz_{i,j}$ is a~$p\times 1$ vector of  dummy covariates that arranges the impact of accident and development year on the claim amounts through model parameters $\bth\in\mathbbm{R}^{p\times 1}$.  The relation~\eqref{muij} defines the \emph{linear predictor} ${\eta}_i=\bz_{i,j}^{\top}\bth$, which together with the link function $g$ fully specifies the \emph{mean structure} $\bmu_i$.

%

Besides the mean structure, 
one needs to specify the \emph{variance} of claim amounts. Assume that the variance of the incremental claim amount $X_{i,j}$ can be expressed as a~\emph{known function} $h$ of its expectations $\mu_{i,j}$:
\begin{equation}\label{ex-var}
\var X_{i,j}=\phi h(\mu_{i,j}),
\end{equation}
where $\phi>0$ is a~\emph{scale} or 
a~\emph{dispersion parameter}. In connection to the GLM, if $X_{i,j}$ followed the Poisson distribution (or the overdispersed Poisson distribution), then $h$ would be an identity, i.e., $h(x)=x$. For the gamma distribution, $h(x)=x^2$, etc. The relation \eqref{ex-var} defines so-called \emph{variance function}.

In the GEE framework, it is \emph{not necessary} to specify the \emph{whole distribution} of the data, because the method is quasi-likelihood based. Only the mean structure and the mean-variance relationship need to be defined. Furthermore, the correlation between the components of $\bX_i$ is modeled using a~\emph{working correlation matrix} $\bC_i(\bm{\vartheta})\in\mathbbm{R}^{(n-i+1)\times (n-i+1)}$, 
which depends only on an~$s\times 1$ vector of unknown parameters $\bm{\vartheta}$, which is the same for all the accident years $i$. Consequently, the \emph{working covariance matrix} of the incremental claims is
\begin{equation}\label{Vi}
\bm{V}_i=\cov \bX_i=\phi\bA_i^{1/2}\bC_i(\bm{\vartheta})\bA_i^{1/2},
\end{equation}
where $\bA_i$ is an~$(n-i+1)\times (n-i+1)$ diagonal matrix with $h(\mu_{i,j})$ as the $j$th diagonal element. The name ``working'' comes from the fact that the structure of $\bC_i$ does not need to be correctly specified. Some commonly used correlation structures are described in Section~\ref{sec_gee_cor}.

To sum up, the GEE framework has two pillars common with the GLM framework (linear predictor and link function). 
However, the \emph{third pillar is different}: 
The GEE approach does not require any specification of the whole distribution for the outcome as this is the case for  the GLM. On contrary, the GEE only assume that the (unknown) distribution belongs to the exponential family of probability distributions and the third pillar consists of specification of the \emph{variance-covariance structure} (variance function and working correlation matrix).

\subsection{Estimation in GEE}\label{sec_gee_est}
The generalized estimating equations are formed via quasi-score vector
\[
\bu(\bth)=\sum_{i=1}^n \bD_i^{\top} \bm{V}_i^{-1}(\bX_i-\bmu_i),
\]
where $\bD_i=\dif \bmu_i/\dif \bth\equiv\left\{\dif\mu_{i,j} /\dif\theta_k \right\}_{j,k=1}^{n-i+i,p}$. For given estimates $(\widehat{\phi},\widehat{\bm{\vartheta}})$ of $(\phi,\bm{\vartheta})$, the estimate of parameter $\bth$ solves the equation $\bu(\widehat{\bth})=\zero$. The parameters $(\phi,\bm{\vartheta})$ are usually estimated by the moment estimates. The fitting algorithm is therefore iterative: updating the estimate of $\bth$ in one step and re-estimating $(\phi,\bm{\vartheta})$ in the second step. 

The procedure yields a~consistent and asymptotically normal estimate of $\bth$ even though the correlation matrix $\bC_i(\bm{\vartheta})$ is misspecified, see~\cite{liang}.  
The empirically corrected variance estimates for $\widehat{\bth}$ can be obtained using the so-called \emph{sandwich estimate}
\begin{equation}\label{sandwichSigma}
\bSigma_{\widehat{\bth}}\equiv\widehat{\cov} \widehat{\bth}=\bB^{-1}(\widehat{\bth}) \bS(\widehat{\bth}) \bB^{-1}(\widehat{\bth}),
\end{equation}
where 
\begin{equation}\label{BS}
\bB=\sum_{i=1}^n \bD_i^{\top} \bm{V}_i^{-1}\bD_i, \quad
 \bS=\sum_{i=1}^n \bD_i^{\top} \bm{V}_i^{-1} (\bX_i-\bmu_i)(\bX_i-\bmu_i)^{\top} \bm{V}_i^{-1}\bD_i
\end{equation}
are evaluated at $\wh{\bth}$. 
The matrix $\bB^{-1}(\widehat{\bth})$ is referred to as a~\emph{model based} estimator of the variance matrix of $\widehat{\bth}$. The estimator $\bSigma_{\widehat{\bth}}$ is consistent for $\cov\widehat{\bth}$ even if the correlation matrix $\bC_i$ is misspecified. However, it can be slightly biased in small samples.

\subsection{Covariance Structure}\label{sec_gee_cor}
Although the GEE method is robust to a misspecification of the correlation structure, selection of the working correlation structure, which is closer to the true one, leads to more efficient estimates of $\bth$. 

There exist several common choices for the working correlation matrix. The simplest case is to assume \emph{uncorrelated} (or~\emph{independent}) incremental claims, i.e., $\bC_i(\bm{\vartheta})=\bI_{n-i+1}=\{\delta_{j,k}\}_{j,k=1}^{n-i+1,n-i+1}$, where $\delta_{j,k}$ symbolizes the Kronecker's delta being $1$ for $j=k$ and $0$ otherwise. The opposite extreme case is an~\emph{unstructured} correlation matrix $\bC_i(\bm{\vartheta})=\{\vartheta_{j,k}\}_{j,k=1}^{n-i+1,n-i+1}$ such that $\vartheta_{j,j}=1$ for $j=1,\ldots,n-1+1$ and $\bC_i(\bm{\vartheta})$ is positive definite. 
As a~compromise to these two extreme cases, one can consider an~\emph{exchangeable} correlation structure
\[
\bC_i(\bm{\vartheta})=\{\delta_{j,k}+(1-\delta_{j,k})\vartheta\}_{j,k=1}^{n-i+1,n-i+1},\quad\bm{\vartheta}=[\vartheta,\ldots,\vartheta]^{\top};
\]
an~\emph{$m$-dependent} correlation structure $\bC_i(\bm{\vartheta})=\{c_{j,k}\}_{j,k=1}^{n-i+1,n-i+1}$,
\[
c_{j,k}=\left\{\begin{array}{lll}
1, & j=k, & \\
\vartheta_{|j-k|}, & 0<|j-k|\leq m, & \bm{\vartheta}=\{\vartheta_l\}_{l=1}^m,\\
0, & |j-k|> m; &
\end{array}\right.
\]
or an~\emph{autoregressive AR(1)} correlation structure
\[
\bC_i(\bm{\vartheta})=\{\vartheta^{|j-k|}\}_{j,k=1}^{n-i+1,n-i+1},\quad\bm{\vartheta}=[\vartheta,\ldots,\vartheta]^{\top}.
\]

\subsection{Application of the GEE to Claims Reserving}\label{sec_gee_claims}
%

%

In the claims reserving, the link function is usually chosen as the logarithm, see, e.g., \cite{wutrich_kniha}. The most common mean structure assumes that
\begin{equation}\label{log-additive}
\log(\mu_{i,j})=\gamma+\alpha_i+\beta_j,
\end{equation}
where $\alpha_i$ stands for the effect of accident year $i$, $\beta_j$ represents the effect of the development year $j$, and $\gamma$ is so-called baseline parameter corresponding a~value for the first accident and development year (taking $\alpha_1=0=\beta_1$). In this case $\bth=[\gamma,\alpha_2,\ldots,\alpha_n,\beta_2,\ldots,\beta_n]$ and
\[
\bz_{i,j}=[1,\delta_{2,i},\ldots,\delta_{n,i},\delta_{2,j},\ldots,\delta_{n,j}]^{\top}.
\]
Another common model, the Hoerl curve with the logarithmic link function, can be coded by design matrix
\[
\bz_{i,j}=[1,\delta_{2,i},\ldots,\delta_{n,i},1\times\delta_{2,j},\ldots,n\times\delta_{n,j},\delta_{2,j}\times\log 2,\ldots,\delta_{n,j}\times\log n]^{\top}
\]
and parameters of interest $\bth=[\gamma,\alpha_2,\ldots,\alpha_n,\beta_2,\ldots,\beta_n,\lambda_2,\ldots,\lambda_n]^{\top}$. Afterwards, $\log(\mu_{i,j})=\gamma+\alpha_i+j\beta_j+\lambda_j\log j$, where again $\alpha_1=\beta_1=\lambda_1=0$. 
For some other possible mean structures in claims reserving, see~\cite{bjorkwall}.


The choice of the variance function is somehow analogous to the specification of the distribution in the GLM.  Hence, suitable variance functions for the claims reserving purposes are: linear (its quasi-score vector corresponds to the score vector of the overdispersed Poisson distribution) or quadratic (gamma distribution). The variance function as a~non-integer power of the mean (multiplied by the scaling parameter) can also be a~practical choice if one realizes the concordance with the Tweedie distribution~\citep{tweedie1984}. This distribution has been recently proven as suitable one for the claims reserving~\citep{wutrich2003} and can be 
considered within the GEE as well.
 
Finally, one needs to choose an appropriate working correlation structure. The most feasible choice 
might be \emph{AR(1)} since the observations within an accident year are \emph{ordered in time} and, in such situations, it is natural that the correlation between two observations decays with their time distance. However, in situations, where the observations are strongly dependent---that is the decay of the correlations is slower than it is in AR(1)---the \emph{exchangeable structure} could be considered as a~good guess as well. Finally, the \emph{independence structure} should always be considered for a~comparison. This approach combined with the sandwich estimate of the covariance matrix of parameter estimates $\widehat{\bth}$ may lead to satisfactory results as well, for small data sets in particular~\citep[Chap.~4]{gee1}.

\section{Model Selection}\label{sec_model}
Similarly as in the GLM setting, two nested models (nested in the mean structure) can be compared using Wald tests, see~\citet[Sec.~4.5.2]{gee1}. A~comparison of two non-nested models in the GLM framework can be based on information criteria as AIC or BIC, see, e.g., \cite{bjorkwall}. However, since the GEE method is only quasi-likelihood based (and not full likelihood), these criteria  cannot be used within the GEE. 
 
\cite{pan} suggested an analogy of the AIC for GEE, namely \emph{quasi-likelihood under the independence model criterion} (QIC). The QIC is defined as
\[
\mathrm{QIC}= -2 Q(\widehat{\bth},\bI) + 2\mathsf{trace}(\widehat{\boldsymbol\Omega}_I(\widehat{\bth}) \bSigma_{\widehat{\bth}}),
\]
where $Q(\cdot,\bI)$ is the quasi-likelihood under working independence model, 
see~\citet[p.~325]{nelder}, and $\widehat{\boldsymbol\Omega}_I(\bth)=\sum_{i=1}^n \bD_i^{\top} \bm{A}_i^{-1}\bD_i$.
A~model with a~\emph{smaller QIC} value indicates a~\emph{better fit} to the data. The QIC equals to AIC (up to a~constant) under the independence in cases when the model implies the proper likelihood. 

\cite{gee1} considered a~modified version of QIC,
\[
\mathrm{QIC}_{HH}= -2 Q(\widehat{\bth},\bI) + 2\mathsf{trace}(\widetilde{\boldsymbol\Omega}_I(\widehat{\bth}(\bI)) \bSigma_{\widehat{\bth}}),
\]
where $\widetilde{\boldsymbol\Omega}_I(\bth)=\sum_{i=1}^n\bD_i^{\top} \bm{A}_i^{-1}\bD_i$ is evaluated at the estimate $\widehat{\bth}(\bI)$ obtained by GEE with the independence working correlation structure. The main advantage of this modification is that  $\mathrm{QIC}_{HH}$  can be  easily computed, because matrices $\widetilde{\boldsymbol\Omega}_I(\widehat{\bth}(\bI))$ and $\bSigma_{\widehat{\bth}}$ are provided by standard software packages for the GEE estimation.

The two criteria can be used for choosing the appropriate mean structure as well as the working correlation matrix.  However, simulations have shown that QIC tends to be more sensitive to changes in the mean structure than changes in the covariance structure, see \cite{hin}. For this reason, \cite{hin} suggested a~\emph{correlation information criterion} (CIC), which improves the performance of QIC  for selecting the appropriate working correlation structure. The CIC is defined as
\[
\mathrm{CIC}=\mathsf{trace}(\widehat{\boldsymbol\Omega}_I(\widehat{\bth}) \bSigma_{\widehat{\bth}}).
\]
Analogously, its modification defined as
\[
\mathrm{CIC}_{HH}=\mathsf{trace}(\widetilde{\boldsymbol\Omega}_I(\widehat{\bth}(\bI))\bSigma_{\widehat{\bth}})
\]
can be used for the comparison of working correlation structures as well.

\section{Mean Square Error of Prediction}\label{sec_mse}
In order to quantify the precision of the estimates and predictions, let us define the \emph{mean square error} (MSE) of prediction for the $i$th claims reserve
\begin{align}
&\mse\left[\widehat{R}_i^{(n)}\right]:=\E\left[\widehat{R}_i^{(n)}-R_i^{(n)}\right]^2=\E\left[\sum_{j=n+2-i}^{n}\left(\widehat{X}_{i,j}^{(n)}-X_{i,j}\right)\right]^2\nonumber\\
&=\sum_{j=n+2-i}^{n}\E\left[\widehat{X}_{i,j}^{(n)}-X_{i,j}\right]^2 +\sum_{\begin{subarray}{c}
j,k=n+2-i\\j\neq k
\end{subarray}}^{n}\E\left[\widehat{X}_{i,j}^{(n)}-X_{i,j}\right]\left[\widehat{X}_{i,k}^{(n)}-X_{i,k}\right],\label{MSEdef}
\end{align}
where $\widehat{X}_{i,j}^{(n)}=
g^{-1}\left(\bz_{i,j}^{\top}\widehat{\bth}\right)$ is the \emph{plug-in prediction} of the incremental claim amounts $X_{i,j}$ based on the GEE estimate $\widehat{\bth}$.

\subsection{Mean Square Error in the GEE}
Our aim is to derive the MSE of prediction for the claims reserves within the GEE framework. Elaborating the expected value from the first sum in~\eqref{MSEdef} yields
\begin{align}
\mse\left[\widehat{X}_{i,j}^{(n)}\right]&:=\E\left[\widehat{X}_{i,j}^{(n)}-X_{i,j}\right]^2=\var\left[\widehat{X}_{i,j}^{(n)}-X_{i,j}\right]+\left(\E\left[\widehat{X}_{i,j}^{(n)}-X_{i,j}\right]\right)^2\nonumber\\
&=\var\widehat{X}_{i,j}^{(n)}-2\cov\left(\widehat{X}_{i,j}^{(n)},X_{i,j}\right)+\var X_{i,j} +\left(\E\widehat{X}_{i,j}^{(n)}-\E X_{i,j}\right)^2.\label{MSEj}
\end{align}
Many authors directly assume the unbiasedness or approximate unbiasedness of estimator $\widehat{X}_{i,j}^{(n)}$ for $\E X_{i,j}$, that is $\E \widehat{X}_{i,j}^{(n)}=\E X_{i,j}$ or $\E\widehat{X}_{i,j}^{(n)}\approx \E X_{i,j}$. See, for instance, \cite{renshaw1994}, \citet[Subsec.~7.1.2]{england2002}, or~\citet[Sec.~3.1]{wutrich_kniha}. Nevertheless, this is neither the case for the GLM nor the GEE, because a~non-linear link function (e.g., logarithm) makes \emph{biased prediction} from approximately unbiased parameter estimates---especially in small samples---due to the non-exchangeability of the expectation operator and the link function. 

The conjecture of the (approximately) unbiased predictor then implies that the MSE of prediction for incremental claims is given as $\mse\left[\widehat{X}_{i,j}^{(n)}\right]\equiv\E\left[\widehat{X}_{i,j}^{(n)}-X_{i,j}\right]^2\approx\var\left[\widehat{X}_{i,j}^{(n)}-X_{i,j}\right]$. That is, the MSE is reduced to the variance of the difference between observation and its prediction. However, the unbiasedness of $\widehat{X}_{i,j}^{(n)}$ can be arguable (or even unrealistic), for smaller samples in particular.

If the prediction is really unbiased and the incremental claim amounts are independent, then the MSE of prediction is equal to the process variance plus the estimation variance, see, e.g., \citet[Sec.~3.1]{wutrich_kniha}. On the other hand, violation of such strict assumptions could provide incorrect MSE of prediction, because it simply ignores the covariance or the squared bias term in~\eqref{MSEj}.

Nevertheless, such simplification cannot be applied in the GEE framework, because \emph{the incremental claim amounts} are \emph{not independent}. 
Hence, the covariance among a~future observation and its predictor in~\eqref{MSEj} is not zero anymore, because the predictor is a~function of the past observations, which are not independent of the future observation. And this needs to be taken into account in the calculation of the MSE.

For $i=2,\ldots,n$ define $\vec{\bX}_i=[X_{i,n+2-i},\ldots,X_{i,n}]^{\top}$ as the vector of the unobserved claim amounts of accident year $i$. Similarly, an arrow above a~vector/matrix stands for its complement for the unobserved data $\{X_{i,j}\}$, $i=2,\ldots,n$ and $j=n+2-i,\ldots,n$ (bottom-right right-angled isosceles triangle). For instance, $\vec{\bmu}_i=\E\vec{\bX}_i$ stands for the expectation of the future claims of accident year $i$, 
$\widehat{\vec{\bX}}_i^{(n)}$ is the prediction of $\vec{\bX}_i$, $\vec{\bD}_i=\partial \vec{\bmu}_i/ \partial \bth$, etc.

%
%

Consider a~first-order stochastic Taylor expansion~\citep[Proposition~6.1.6]{bd}, for the residual vector $\vec{\br}_i:=\vec{\bX}_i-\widehat{\vec{\bX}}_i^{(n)}$ around $\bth$. It gives  
\begin{equation}\label{taylor}
\vec{\br}_i=\vec{\be}_i+\left[\frac{\dif\vec{\be}_i}{\dif\bth}\right](\widehat{\bth}-\bth)+o_{\P}(\|\widehat{\bth}-\bth\|),
\end{equation}
where $\vec{\be}_i\equiv\vec{\be}_i(\bth)=\vec{\bX}_i-\vec{\bmu}_i$. Notice that $\dif\vec{\be}_i/\dif\bth=-\vec{\bD}_i$. Previous linearization is reasonable, because under the regularity conditions for quasi-likelihood estimation in the GEE framework postulated by~\cite{white1982} and~\citet[Sec.~5.2]{gee2}, the quasi-likelihood GEE estimate $\widehat{\bth}$ is strongly consistent for the parameter $\bth$.

The MSE of $\widehat{\vec{\bX}}_i^{(n)}$ can be calculated using residuals $\vec{\br}_i$ and \eqref{taylor} as
\begin{align}
\mse\left[\widehat{\vec{\bX}}_i^{(n)}\right]=\E\left[\vec{\br}_i\vec{\br}_i^{\top}\right]&\approx\E\left[\vec{\be}_i\vec{\be}_i^{\top}\right]-\E\left[\vec{\be}_i(\widehat{\bth}-\bth)^{\top}\vec{\bD}_i^{\top}\right]-\E\left[\vec{\bD}_i(\widehat{\bth}-\bth)\vec{\be}_i^{\top}\right]\nonumber\\
&\quad +\E\left[\vec{\bD}_i(\widehat{\bth}-\bth)(\widehat{\bth}-\bth)^{\top}\vec{\bD}_i^{\top}\right].\label{MSErr}
\end{align}
The Taylor expansion applied on $\bu(\widehat{\bth})$ around $\bth$ together with the chain rule provide a first-order approximation
\[
\widehat{\bth}-\bth\approx\left(\sum_{l=1}^n\bD_l^{\top}\bV_l^{-1}\bD_l\right)^{-1}\sum_{j=1}^n\bD_j^{\top}\bV_j^{-1}\be_j.
\]
For a~detailed derivation see~\citet[Sec.~5.2]{gee2}. This approximation together with the independence of accident years $i$ and $j$, $i\ne j$, imply that~\eqref{MSErr} can be further expressed as
\begin{align}
\mse\left[\widehat{\vec{\bX}}_i^{(n)}\right]&\approx\cov\vec{\bX}_i-\cov\left(\vec{\bX}_i,\bX_i\right)\bH_{ii}^{\top}-\bH_{ii}\cov\left(\bX_i,\vec{\bX}_i\right)\nonumber\\
&\quad+\sum_{j=1}^n\bH_{ij}\cov\bX_j\bH_{ij}^{\top}\label{MSEH}\\
&=\cov\vec{\bX}_i-2\cov\left(\vec{\bX}_i,\bX_i\right)\bH_{ii}^{\top}+\vec{\bD}_i\bB^{-1}[\E\bS]\bB^{-1}\vec{\bD}_i^{\top},\label{MSEHs}
\end{align}
where $\bH_{ij}=\vec{\bD}_i\left(\sum_{l=1}^n\bD_l^{\top}\bV_l^{-1}\bD_l\right)^{-1}\bD_j^{\top}\bV_j^{-1}$ and the matrices $\bB$ and $\bS$ are defined in \eqref{BS}.

\subsection{Estimate for the MSE of Prediction}
One of the main goals is to estimate the theoretical MSE of prediction for claims reserves. This means to find a~proper estimate for the left hand side of approximation~\eqref{MSEH}. Indeed, comparing relations~\eqref{MSEdef} and~\eqref{MSErr} gives
\[
\mse\left[\widehat{R}_i^{(n)}\right]=\underbrace{[1,\ldots,1]}_{(i-1)\times 1}\mse\left[\widehat{\vec{\bX}}_i^{(n)}\right][1,\ldots,1]^{\top}.
\]

The core problem lies in the estimation of the covariances in~\eqref{MSEHs}. The remaining terms from~\eqref{MSEHs} can be  estimated straightforwardly using the plug-in estimates, i.e.,
\begin{alignat*}{3}
\widehat{\bA}_i&:=\bA_i(\widehat{\bth}), &\quad \widehat{\bC}_i&:=\bC_i(\widehat{\bvartheta}), &\quad \widehat{\bV}_i&:=\widehat{\phi}\widehat{\bA}_i^{1/2}\widehat{\bC}_i\widehat{\bA}_i^{1/2},\\\widehat{\bD}_i&:=\bD_i(\widehat{\bth}), &\quad \widehat{\vec{\bD}}_i&:=\vec{\bD}_i(\widehat{\bth}). &\quad
\end{alignat*}
The covariance of the observed incremental claim amounts can be estimated as suggested by~\cite{liang}:
\[
\widehat{\cov}\bX_i=(\bX_i-\widehat{\bX}_i)(\bX_i-\widehat{\bX}_i)^{\top}.
\]
It follows from~\eqref{sandwichSigma}, that the last term of~\eqref{MSEHs} can be estimated by $\widehat{\vec{\bD}}_i\bSigma_{\widehat{\bth}}\widehat{\vec{\bD}}_i^{\top}$. Furthermore, the variance structure in~\eqref{Vi} implies that the covariance of the unobserved (future) incremental claim amounts may be estimated 
as
\[
\widehat{\cov}\vec{\bX}_i=\widehat{\phi}\widehat{\vec{\bA}}_i^{1/2}\widehat{\vec{\bC}}_i\widehat{\vec{\bA}}_i^{1/2},
\]
where $\widehat{\vec{\bA}}_i=diag\{h(\mu_{i,j}(\widehat{\bth})),\,j=n+2-i,\ldots,n\}$ and $\widehat{\vec{\bC}}_i 
=\bC_{n+2-i}(\widehat{\bvartheta})$ for the standard correlation structures as AR$(1)$, MA$(1)$, independence, or exchangeable (i.e., correlation structures with the translation symmetry property). If a~different correlation structure is used, then the future correlations $\vec{\bC}_i$ have to be predefined in advance and estimated according to that. 

Another way how to look at the covariances from~\eqref{MSEH} is to consider a~joint vector of the past and future incremental claim amounts for a~particular accident year. Hence,
\begin{equation}\label{covXX}
\cov[\bX_i^{\top},\vec{\bX}_i^{\top}]^{\top}=\left[\begin{array}{cc}
\cov\bX_i & \cov\left(\bX_i,\vec{\bX}_i\right) \\
\cov\left(\vec{\bX}_i,\bX_i\right) & \cov\vec{\bX}_i
\end{array}\right]=\phi\tilde{\bA}_i^{1/2}\tilde{\bC}_i\tilde{\bA}_i^{1/2},
\end{equation}
where $\tilde{\bA}_i=diag\{h(\mu_{i,j}(\bth)),\,j=1,\ldots,n\}\equiv\,diag\{diag(\bA_i),diag(\vec{\bA}_i)\}$, i.e., joint diagonals from $\bA_i$ and $\vec{\bA}_i$ are placed on the diagonal of matrix $\tilde{\bA}_i$.
The correlation matrix 
$\tilde{\bC}_i$ is an~extension of the original correlation matrix $\bC_i$ for the standard correlation structures as above, or needs to be known in advance. 


Henceforth, the estimate of covariance among the past and future incremental claim amounts can easily be taken from~\eqref{covXX}, i.e.,
\[
\widehat{\cov}\left(\vec{\bX}_i,\bX_i\right)=\widehat{\phi}\widehat{\vec{\bA}}_i^{1/2}\widehat{\bar{\bC}}_i\widehat{\bA}_i^{1/2},
\]
where $\bar{\bC}_i$ is a lower-left segment of the correlation matrix $\tilde{\bC}_i$ corresponding to $\cov\left(\vec{\bX}_i,\bX_i\right)$, i.e., $\bar{\bC}_i=\{\tilde{\bC}_{i;j,k}\}_{j=n+2-i,k=1}^{n,n+1-i}$.

In order to calculate the estimate for the total claims reserve $R^{(n)}$, one just needs to sum up the claims reserve's estimates for each accident year due to the fact that the claim amounts in different accident years are independent. Hence,
\begin{subequations}\label{MSEest}
\begin{align}
\widehat{\mse}\left[\widehat{R}^{(n)}\right]&=
\sum_{i=2}^{n}\underbrace{[1,\ldots,1]}_{(i-1)\times 1}
\widehat{\mse}\left[\widehat{\vec{\bX}}_i^{(n)}\right]
\left[\begin{array}{c}
1\\
\vdots\\
1
\end{array}
\right],\\
\widehat{\mse}\left[\widehat{\vec{\bX}}_i^{(n)}\right]&=\widehat{\phi}\widehat{\vec{\bA}}_i^{1/2}\widehat{\vec{\bC}}_i\widehat{\vec{\bA}}_i^{1/2}
-2\widehat{\phi}\widehat{\vec{\bA}}_i^{1/2}\widehat{\bar{\bC}}_i\widehat{\bA}_i^{1/2}\widehat{\bH}_{ii}^{\top}
+\widehat{\vec{\bD}}_i\bSigma_{\widehat{\bth}}\widehat{\vec{\bD}}_i^{\top},\\
\intertext{and}
\widehat{\bH}_{ij}&=\widehat{\vec{\bD}}_i\left(\sum_{l=1}^n\widehat{\bD}_l^{\top}\widehat{\bV}_l^{-1}\widehat{\bD}_l\right)^{-1}\widehat{\bD}_j^{\top}\widehat{\bV}_j^{-1}.
\end{align}
\end{subequations}


\section{Real Data Illustration}\label{sec_data}
The real data analyses are conducted in R~program~\citep{Rko} using functions {\sf geeglm} and {\sf geese} from package {\sf geepack} \citep{geepack}. In all the presented GEE models for claim triangles, additive accident-development year mean structure~\eqref{log-additive} with the logarithmic link function is used. The independence, exchangeable, and AR$(1)$ correlation structures are considered. The variance function is chosen as linear or quadratic (i.e., corresponding to an overdispersed Poisson or a gamma model in the GLM). This means that for each data set, six different models with the same mean structure are fitted. Competing models are compared using $\text{QIC}_{HH}$ and $\text{CIC}_{HH}$ as these criteria can be easily obtained from the output of the function {\sf geeglm}.

\subsection{Data by~\cite{taylor_ashe}}
Firstly, we illustrate the proposed method on a~data set from~\cite{taylor_ashe}. Here, $n=10$ accident years are available. The Pearson residuals of the classical GLM suggest that there might be some small or moderate correlation of the incremental claims within the same accident year (for the first and second development years, we get $-0.22$ for the overdispersed Poisson and $-0.29$ for the gamma model). Hence, the application of the GEE might be suitable.

The estimated reserves for the six competing models are listed in Table~\ref{tab:comparisonR} (bottom half of the table). Table~\ref{tab:comparisonR} also contains reserve estimates from other well-known reserving methods for a comparison. All these reserve estimates are taken from Table~1 in~\cite{EV1999}, where a~brief description of all the methods is provided as well.

It should be noticed that the GEE model with independence correlation structure and quadratic variance function provides exactly the same reserve estimates as the GLM gamma model. And similarly, the GEE with independence correlation structure and linear variance function gives the same reserve estimates as the GLM overdispersed Poisson model. Indeed, the GLM are sometimes called \emph{Independent Estimating Equations} \citep{liang} and in that case the quasi-likelihood and full-likelihood approach coincide. However, as it will be seen later on, there are noticeable differences in the models in terms of the MSE of prediction (partially caused by the use of the robust sandwich covariance matrix estimate in the GEE).

\begin{landscape}
\begin{table}[!ht]
\begin{center}
{\footnotesize\begin{tabular}{lrrrrrrr}
\hline
Accident & Chain & Poisson & Gamma & \cite{mack1991} & \cite{verrall1991} & Renshaw/ & Zehnwirth\\
year & ladder & GLM & GLM & & & Christofides &\\
\hline
$i=2$ & 95 & 95 & 93 & 93 & 96 & 111 & 109\\
$i=3$ & 470 & 470 & 447 & 447 & 439 & 482 & 473\\
$i=4$ & 710 & 710 & 611 & 611 & 608 & 661 & 648\\
$i=5$ & 985 & 985 & 992 & 992 & 1\,011 & 1\,091 & 1\,069\\
$i=6$ & 1\,419 & 1\,419 & 1\,453 & 1\,423 & 1\,423 & 1\,531 & 1\,500\\
$i=7$ & 2\,178 & 2\,178 & 2\,186 & 2\,186 & 2\,150 & 2\,311 & 2\,265\\
$i=8$ & 3\,920 & 3\,920 & 3\,665 & 3\,665 & 3\,529 & 3\,807 & 3\,831\\
$i=9$ & 4\,279 & 4\,279 & 4\,122 & 4\,122 & 4\,056 & 4\,452 & 4\,364\\
$i=10$ & 4\,626 & 4\,626 & 4\,516 & 4\,516 & 4\,340 & 5\,066 & 4\,965\\
\hline
Total & 18\,681 & 18\,681 & 18\,085 & 18\,085 & 17\,652 & 19\,512 & 19\,124\\
\hline
 &  & GEE Ind & GEE Ind & GEE Exch & GEE Exch & GEE AR(1) & GEE AR(1)\\
 &  & linear & quadratic & linear & quadratic & linear & quadratic\\
\hline
$i=2$ &  & 95 & 93 & 100 & 93 & 85 & 90\\
$i=3$ &  & 470 & 447 & 473 & 447 & 443 & 431\\
$i=4$ &  & 710 & 611 & 683 & 611 & 706 & 618\\
$i=5$ &  & 985 & 992 & 1\,014 & 992 & 970 & 968\\
$i=6$ &  & 1\,419 & 1\,453 & 1\,445 & 1\,453 & 1\,382 & 1\,412\\
$i=7$ &  & 2\,178 & 2\,186 & 2\,194 & 2\,186 & 2\,166 & 2\,167\\
$i=8$ &  & 3\,920 & 3\,665 & 3\,891 & 3\,665 & 3\,809 & 3\,611\\
$i=9$ &  & 4\,279 & 4\,122 & 4\,279 & 4\,122 & 4\,221 & 4\,090\\
$i=10$ &  & 4\,626 & 4\,516 & 4\,631 & 4\,516 & 4\,585 & 4\,483\\
\hline
Total &  & 18\,681 & 18\,086 & 18\,710 & 18\,086 & 18\,367 & 17\,870\\
\hline
\end{tabular}}
\end{center}
\caption{Reserve estimates (in thousands) for~\cite{taylor_ashe} data based on various reserving methods.}
\label{tab:comparisonR}
\end{table}
\end{landscape}

In addition, the reserve estimates for the GEE quadratic independence model and the quadratic AR(1) are very similar, but not identical (differences in hundreds or tens).

$\text{QIC}_{HH}$ and $\text{CIC}_{HH}$ criteria for the comparison of the GEE models are listed in Table~\ref{taylor_qic}. Both criteria for the linear and quadratic variance function favor the independence working correlation structure. Hence, for this data set, it seems that estimates obtained by this working correlation structure with the robust standard errors obtained from the sandwich estimator $\bSigma_{\widehat{\bth}}$ might be used for the predictions. However, as it can be seen from Table~\ref{taylor_qic}, the AR(1) and exchangeable structure could be reasonable as well, because the differences in the values of criteria are rather small.


\begin{table}[!ht]
\begin{center}
\begin{tabular}{lrrrr}
\hline
Covariance & \multicolumn{2}{c}{Linear variance function} & \multicolumn{2}{c}{Quadratic variance function} \\
structure & $\QIC_{HH}$ & $\CIC_{HH}$ & $\QIC_{HH}$ & $\CIC_{HH}$\\
\hline
Independence & $-$857\,098\,696 & 9.48 & 1\,583.20 & 10.66 \\
Exchangeable & $-$857\,080\,756 & 9.58 & 1\,583.20 & 10.66\\
AR(1) & $-$857\,086\,975 & 9.68 & 1\,583.58 & 10.85\\
\hline
\end{tabular}
\end{center}
\caption{$\QIC_{HH}$ and $\CIC_{HH}$ criteria for~\cite{taylor_ashe} data.}
\label{taylor_qic}
\end{table}

%
%

Looking at Table~\ref{tab:comparisonR}, one can conclude, that the reserve estimates (for each accident year as well as in total) are quite comparable, with some small differences. This holds for all the proposed GEE models as well as for the other reserving methods compared by~\cite{EV1999}.

It is well known that a~decision for the suitable model should be based on goodness of fit methods, because they measure discrepancy between the model and the data. The choice of the final model should not be made according to the estimate of MSE of prediction, because less variable prediction does not have to straightforwardly imply better model fit to data. Nevertheless, we have compared the models from Table~\ref{tab:comparisonR} in terms of their precision, e.g., the MSE of predictions. In Table~\ref{tab:comparisonMSE}, the estimated MSEs of prediction are summarized. 

\begin{landscape}
\begin{table}[!ht]
\begin{center}
{\footnotesize\begin{tabular}{lrrrrrrr}
\hline
Accident & \cite{mack} & Poisson & Gamma & \cite{mack1991} & \cite{verrall1991} & Renshaw/ & Zehnwirth\\
year & distribution free & GLM & GLM & & & Christofides &\\
\hline
$i=2$ & 80 & 116 & 48 & 40(49) & 49 & 54 & 49\\
$i=3$ & 26 & 46 & 36 & 30(37) & 37 & 39 & 35\\
$i=4$ & 19 & 37 & 29 & 24(30) & 30 & 32 & 29\\
$i=5$ & 27 & 31 & 26 & 21(26) & 27 & 28 & 25\\
$i=6$ & 29 & 26 & 24 & 20(25) & 25 & 26 & 24\\
$i=7$ & 26 & 23 & 24 & 20(25) & 25 & 26 & 24\\
$i=8$ & 22 & 20 & 26 & 21(26) & 27 & 28 & 26\\
$i=9$ & 23 & 24 & 29 & 24(30) & 30 & 31 & 30\\
$i=10$ & 29 & 43 & 37 & 31(38) & 38 & 40 & 39\\
\hline
Total & 13 & 16 & 15 & $-$ & 15 & 16 & 16\\
\hline
 & Bootstrap CL & GEE Ind & GEE Ind & GEE Exch & GEE Exch & GEE AR(1) & GEE AR(1)\\
 & (independence) & linear & quadratic & linear & quadratic & linear & quadratic\\
\hline
$i=2$ & 80 & 60 & 26 & 63 & 30 & 60 & 26\\
$i=3$ & 26 & 28 & 24 & 32 & 28 & 24 & 23\\
$i=4$ & 19 & 24 & 20 & 27 & 24 & 19 & 18\\
$i=5$ & 27 & 23 & 23 & 26 & 26 & 19 & 21\\
$i=6$ & 29 & 17 & 15 & 20 & 19 & 14 & 13\\
$i=7$ & 26 & 15 & 15 & 18 & 19 & 12 & 13\\
$i=8$ & 22 & 10 & 13 & 13 & 17 & 9 & 12\\
$i=9$ & 23 & 11 & 13 & 14 & 17 & 10 & 12\\
$i=10$ & 29 & 11 & 13 & 17 & 17 & 13 & 14\\
\hline
Total & 13 & 5.1 & 5.6 & 6.9 & 7.5 & 4.8 & 5.5\\
\hline
\end{tabular}}
\end{center}
\caption{Estimated MSE of prediction as $\%$ of reserve estimate for~\cite{taylor_ashe} data from various reserving methods.}
\label{tab:comparisonMSE}
\end{table}
\end{landscape}

Recall that the GEE approach can be considered as a~distribution free, the estimate of the MSE of prediction derived in Section~\ref{sec_mse} does not neglect the bias of prediction, and that the dependencies are allowed within each accident year. Despite these facts, the estimated MSEs of prediction for the GEE models are \emph{remarkably smaller} than in case of other mentioned well-known reserving models (see the relative MSE of prediction in percentages in Table~\ref{tab:comparisonMSE}). 
One of the possible reasons is the form of the MSE's estimate~\eqref{MSEest}, which is derived in a~different, probably more efficient way compared to the other MSE's estimates and which incorporates the robust and consistent properties of the covariance sandwich estimator~\eqref{sandwichSigma}. The other reason can be a~very flexible framework of the GEE.

%
%
%


Another remark involves the estimated MSE of prediction based on the bootstrap approach, which is also compared above to the estimated MSEs from GEE. Bootstrapping residuals independently assumes independent observations. Probable reason for less precise prediction is invalid assumption for the classical bootstrap that the residuals are independent. Henceforth, independent resampling of residuals should be replaced by a~proper cluster bootstrap or block bootstrap. Moreover, non-parametric bootstrap consistency for chain ladder has not been proved yet. It still remains questionable, whether this resampling approach works for all types of triangles (and under which conditions). Only the sufficient and necessary conditions for the consistency of development factors in the chain ladder model has been shown recently, cf.~\cite{PH2012}.

Note that the smallest estimated MSE of prediction within the GEE models (and among all the shown models as well) is for the AR(1) covariance structure with linear variance function ($4.8$). Similarly, the quadratic AR(1) GEE model has smaller MSE of prediction than the quadratic independent GEE one ($5.5<5.6$). However, as already discussed, the model selection criteria slightly favor the independence model. The final decision could be therefore based on some other model diagnostics (e.g., residuals) as well.



\subsection{Data by~\cite{Zehnwirth}}
The second analyzed data come from~\cite{Zehnwirth}. Here, data for $n=11$ accident years are available. Again, residuals from the classical gamma GLM model indicate dependence between claims of the same accident year (correlation $0.596$ for the first and second development year) and, thus, the GEE approach might be appropriate. 

For this data set, the quadratic variance function seems to be more suitable than the linear one. However, for the sake of completeness both variance functions are considered and six different models are fitted. The obtained estimated total outstanding reserves together with their relative prediction errors are listed in Table~\ref{abc}.

\begin{table}[!ht]
\begin{center}
\begin{tabular}{lrrrrrrr}
\hline
 & Ind & Ind & Exch & Exch & AR(1) & AR(1)\\
 & linear & quadratic & linear & quadratic & linear & quadratic\\
\hline
Reserves & 5\,278 & 5\,238 & 5\,258 & 5\,238 & 5\,311 & 5\,269\\
MSE $[\%]$ & 1.90 & 2.14 & 2.00 & 2.61 & 1.95 & 1.96\\
\hline
\end{tabular}
\end{center}
\caption{Total reserve estimates (in thousands) and estimated MSE of prediction as percentages of reserve estimate for~\cite{Zehnwirth} data from various GEE models.}
\label{abc}
\end{table}

$\text{QIC}_{HH}$ and $\text{CIC}_{HH}$ criteria 
are listed in Table~\ref{ABC_qic}. For the linear variance function, criterion $\QIC_{HH}$ is minimal for the independence working structure. On the other hand, $\CIC_{HH}$ is minimal for the AR(1) correlation structure. In case of the quadratic variance function, both criteria favor the AR(1) correlation structure. Hence, the AR(1) dependence structure combined with the quadratic variance function could be the best choice. Note that the smallest MSE of prediction for reserves is in the case of the independence correlation structure with the linear variance function. However, the MSE of the quadratic AR(1) is comparable.

\begin{table}[!ht]
\begin{center}
\begin{tabular}{lrrrr}
\hline
Covariance & \multicolumn{2}{c}{Linear variance function} & \multicolumn{2}{c}{Quadratic variance function} \\
structure & $\QIC_{HH}$ & $\CIC_{HH}$ & $\QIC_{HH}$ & $\CIC_{HH}$\\
\hline
Independence & $-$230\,052\,223 & 10.21 & 1\,682.24 & 11.24\\
Exchangeable & $-$230\,051\,487 & 10.53 & 1\,682.24 & 11.24\\
AR(1) & $-$230\,052\,055 & 9.92 & 1\,681.86 & 11.05\\
\hline
\end{tabular}
\end{center}
\caption{$\QIC_{HH}$ and $\CIC_{HH}$ criteria for~\cite{Zehnwirth} data.}
\label{ABC_qic}
\end{table}

\section{Conclusions and Discussion}
This paper proposes the \emph{GEE modeling technique} as a~suitable stochastic method for claims reserving. Classical stochastic methods usually assume independent claim amounts. If this assumption is violated, then these techniques can provide incorrect and misleading inference. In contrast to this, the GEE approach \emph{enables modeling dependencies} between claim amounts of the development years within each accident year. These dependencies are modeled via a~working correlation matrix. 
Even if this correlation structure is misspecified, the estimates for the claims reserves are still valid and consistent. 
A~correctly specified dependence structure improves the efficiency of the procedure.
On the top of that, the GEE do not require any specific distributional assumptions on the claim amounts.

Model \emph{selection criteria} are available for the GEE and, thus, the competing models can be compared directly by one number, which is often a~practical advantage. However, the
whole model fit cannot be simply characterized by just one number. These criteria should be seen as one of many factors, which could be taken into account in the model selection. For instance, an inspection of residuals should be an~indispensable part of the reserve estimation process. Their diagnostics give insight into the goodness of fit and possible violations of the model assumptions (e.g., mean-variance relationship). It should be also noted that stochastic models from different classes (having different formulation and assumptions) cannot be directly compared by one common criterion. 

The dependencies were considered only within each accident year (origin year clusters). Reasonable argumentation can lead into modeling dependencies in a~diagonal way in the claim triangles. In this case, each calendar year would form a~cluster. A~different notation, than the one used in Section~\ref{sec_gee}, would be needed, but the principles remain the same.


An~\emph{estimate for the mean square error of prediction} for the claims reserves is derived in a~non-traditional way. 
It incorporates the sandwich (robust) covariance matrix estimate~\eqref{sandwichSigma}, it does not neglect the bias of prediction, and it does not ignore dependencies between claim amounts. The performance of this estimate is surprising, as shown in Section~\ref{sec_data}. 
The source of the increase in precision is not in the estimation of the process variance, but it is hidden in the estimation variance, or better to say, in the MSE of the estimate, because we do not ignore the estimate's bias. The classical naive empirical estimates of the MSE 
are less efficient than the one proposed in this paper.

As a~bonus, an alternative and more precise estimate of the MSE of prediction for the GLM is introduced. Indeed, the GEE model with independence correlation structure provides exactly the same estimates as the GLM with a~suitable distribution, which has to correspond to the variance function from the GEE.


%
%
%

Another way how to calculate the MSE of prediction for claims reserves might be a~\emph{cluster bootstrap in GEE} \citep{cheng2013}. This approach would provide more than an~estimate of the MSE, it could provide an~estimate of the whole distribution for the reserves. On the other hand, the cluster bootstrapping generally requires a~data set with more observations than we usually possess in claim triangles.

\section*{Acknowledgments}
The research of \v{S}\'{a}rka Hudecov\'{a} on this paper was supported by the Czech Science Foundation project ``DYME – Dynamic Models in Economics'' No.~P402/12/G097. The work of Michal Pe\v{s}ta on this paper was funded by the Czech Science Foundation project GA\v{C}R No.~P201/13/12994P.

\bibliographystyle{elsarticle-harv}
\bibliography{claims_lit}







\end{document}